\documentclass[final,5p,times,twocolumn]{elsarticle}

\usepackage{graphicx}

\usepackage{amsmath,amssymb,amsbsy,amstext}
\usepackage{soul}
\usepackage{rotating}

\begin{document}

\begin{frontmatter}



\title{Analysis of environmental influences in nuclear half-life measurements exhibiting time-dependent decay rates}


\author[NE,PHYS]{Jere H. Jenkins\corref{cor1}}
\ead{jere@purdue.edu}
\cortext[cor1]{Corresponding author}
\address[NE]{School of Nuclear Engineering, Purdue University, West Lafayette, IN  47907, USA}
\address[PHYS]{Physics Department, Purdue University, West Lafayette, IN 47907, USA}

\author[Mayo]{Daniel W. Mundy}
\address[Mayo]{Department of Radiation Oncology Physics, Mayo Clinic, Rochester, MN 55905, USA}

\author[PHYS]{Ephraim Fischbach}

\begin{abstract}
In a recent series of papers evidence has been presented for correlations between solar activity and nuclear
decay rates. This includes an apparent correlation between Earth-Sun distance and data taken at 
Brookhaven National Laboratory (BNL), and at the Physikalisch-Technische Bundesanstalt (PTB). Although these correlations could
arise from a direct interaction between the decaying nuclei and some particles or fields emanating from the
Sun, they could also represent an ``environmental'' effect arising from a seasonal variation of the sensitivities
of the BNL and PTB detectors due to changes in temperature, relative humidity, background radiation, etc. In this
paper, we present a detailed analysis of the responses of the detectors actually used in the BNL and PTB experiments,
and show that sensitivities to seasonal variations in the respective detectors are likely too small 
to produce the observed fluctuations. 
\end{abstract}


\end{frontmatter}


\section{Introduction}
In a recent series of papers \citep{Jenkins09Fapp,Jenkins09Capp,Fischbach09} 
evidence has been presented suggesting a correlation between nuclear decay rates and solar activity. 
This evidence came from analyses of data from three independent sources. The first was data taken at Purdue University during a period which included the solar flare 
of 2006 December 13, which exhibited a statistically significant dip in the counting rate of 
$^{54}$Mn coincident in time with the solar flare \citep{Jenkins09Fapp}. 
The second was data from a measurement of the half-life (T$_{1/2}$) of $^{32}$Si at 
Brookhaven National Laboratory (BNL) \citep{Alburger86}, and the third data set was an extended study of detector stability via measurements of the decay rates of $^{152}$Eu, $^{154}$Eu, and $^{226}$Ra at the Physikalisch-Technische Bundesanstalt (PTB) in 
Germany \citep{Siegert98,Schrader06}. The BNL and PTB data revealed periodic variations in measured decay rates 
which approximately correlated with Earth-Sun distance. Taken together the data 
of Refs. \citep{Jenkins09Fapp,Jenkins09Capp,Fischbach09,Alburger86,Siegert98,Schrader06} raise the possibility that nuclear 
decays are being influenced by the Sun through some as yet unknown 
mechanism, perhaps involving solar neutrinos. 

The $^{54}$Mn data presented in Ref. \citep{Jenkins09Fapp} were a series of successive four-hour counts 
taken over the course of two months as part of a half-life measurement. The selection of a four-hour 
count time allowed for a time resolution capable of detecting the solar flare, 
which lasted approximately 43 minutes \citep{Verkho08}. Due to the short counting period, the counts were 
presumably insensitive to typical seasonal environmental effects, such as temperature, pressure, background, etc.

However, the same cannot be said of the data of Refs. \citep{Alburger86,Siegert98,Schrader06} 
since the annually varying decay rates reported in these references could in principle 
arise from a number of conventional systematic influences on the detectors employed in these 
experiments rather than from a modulation of the decay rates themselves. The object 
of the present paper is to analyze the BNL and PTB data in detail in an effort to 
disentangle environmental influences on the detector systems used in each experiment 
from a possible contribution originating from the Sun, which could affect the intrinsic decay rates.
The detector system used in the $^{54}$Mn data set associated with the 2006 December 13 solar flare
will also be discussed.

Although there are many mechanisms through which experimental detectors can be influenced 
by environmental factors, a useful starting point for the present analysis is the recent 
paper by Semkow, et al. \citep{Semkow09} in which 
the authors attempt to explain both the BNL \citep{Alburger86,Harbottle73} and PTB \citep{Siegert98,Schrader06,Schrader07} 
data in terms of seasonal temperature variations. In what follows we will analyze these 
experiments in detail and show explicitly why the mechanisms proposed by Semkow et al. to 
explain the BNL/PTB data are not likely to be correct. This analysis can thus serve as a template 
for similar analyses of other past and future experiments on half-life determinations to evaluate
possible seasonal environmental influences.

\section{The PTB Ionization Chamber Detector}

In the course of experiments at the PTB \citep{Siegert98} which were intended 
to evaluate the stability of detector systems employed in the measurement of half-lives, 
e.g. $^{152}$Eu (T$_{1/2}$=4936.6(20)d), the authors reported an oscillation in the decay rate 
of the long-lived standard, $^{226}$Ra (T$_{1/2}$=1600(7)y) used in the measurements. These 
fluctuations were attributed by the authors to a ``discharge effect on the charge collecting 
capacitor, the cables and insulator to the ionization chamber electrode caused by background 
radioactivity such as radon and daughters which are known to show seasonal concentration 
changes \citep{Siegert98}''  . It was also proposed 
elsewhere \citep{Semkow09} that the variations in the measured activity of 
the $^{226}$Ra source were caused by variations of the density of the argon working gas 
within the detector chamber itself, caused by seasonal changes in the ambient temperature. 
This and other potential systematic effects, such as changes in ambient pressure, humidity, 
background radiation or electronic fluctuations, are indeed possible explanations for 
the observed variations and should be examined.  An alternative proposal to these 
explanations, as suggested in Ref. \citep{Jenkins09Capp}, was that the oscillations 
of the $^{226}$Ra measurements were not caused by local perturbations to the detector system, 
but were due to the influence of fields or particles such as neutrinos emanating from the 
Sun.  In what follows, we will examine possible environmental influences on the detector 
system, and determine which of these, if any, could be responsible for the fluctuations 
in the $^{226}$Ra measurements.

To evaluate which environmental influences may have caused the fluctuations in the 
detector, it is helpful to understand how the detector utilized in the PTB experiments works.  The measurements 
were taken with a high-pressure 4$\pi\gamma$ ionization chamber, Model IG12, manufactured 
by 20th Century (now Centronic, Ltd.), which was filled with argon to 20 atm (2 MPa) 
by the manufacturer, then welded shut prior to delivery to the user (Trevor Alloway, 
Centronic Ltd., private communication, 2009). Therefore, the number of atoms of the 
working gas, argon, within the detector is constant throughout the lifetime of the 
detector, irrespective of the external temperature, pressure or relative humidity. 

The ionization chamber for the PTB experiments was operated in a ``current'' 
mode, as is typical with ionization chambers in this application. Here, the output current 
of the detector is generated by the collection of electron-ion pairs created by the 
ionization of the argon gas inside. These electron-ion pairs arise from photon interactions within the gas, and from the 
interaction of photoelectrons and Compton electrons generated by photon interactions 
in the wall of the re-entrant tube, which then enter the working gas also causing ionization.  
Ionization chambers operate at sufficiently high bias voltages to prevent recombination 
of the electron-ion pairs generated by the interactions of photons, photoelectrons, and 
Compton electrons in the gas. In the case of the PTB measurements, that voltage was 
500V. The electrodes (anode and cathode) of the detector are contained within the 
chamber of the detector itself, and these collect the electrons and ions created via the radiation interactions.  The 
current output is then the result of the of electrons collected at the anode, which 
is $\sim1.44\times10^{-8} \rm{Amp\cdot{}Roentgen^{-1}\cdot{}hr^{-1}}$ (http://www.centronic.co.uk/ionisation.htm, 2009). 
One obvious systematic variable that can be quickly examined is variation in the bias 
voltage. However, ionization chambers are known for their insensitivity to applied bias 
voltage changes, since the ionization current is essentially independent of the 
applied voltage in the ion chamber region \citep{Knoll00}.  
This is due to the fact that, as pointed out above, the bias voltage 
applied in an ionization chamber is chosen to be sufficiently high to prevent 
appreciable recombination of the generated electron-ion pairs, while at the same 
time not providing enough energy to cause multiplication of electron-ion pairs within 
the detector. It follows that the output current from the detector will be based 
solely on the number of electron-ion pairs generated by the incident radiation and 
collected at the electrodes within.

Since it is the ionization of the argon gas within the detector that provides the measure of the 
energy deposited within by the incident radiation, and thus the activity of the sources being 
analyzed, it follows that changes in the properties of the gas within the detector, such as the 
atom density, could affect the measurements. However, since the number of moles ($n$) of argon gas 
within the PTB detector is constant, the density $n/V=\rho$ of the gas will also be constant, unless the 
volume $(V)$ of the detector changes. Using the ideal 
gas law, $PV=nRT$, where $R$ is the gas constant, $T$ is the temperature, and $P$ is the pressure, 
it follows that if the volume is also 
constant (a non-constant volume will be addressed below) then the detector system is described 
by $P_{1}/T_{1}=P_{2}/T_{2}$, where $1$ and $2$ denote the initial and final conditions, respectively.  
Hence, the only effect of a change in room temperature is to change the pressure inside the detector.
And, since the ionization potential 
of argon is known to be a constant over all pressures from 1 to 200 bar \citep{Found20}, it follows that
there should be no change in the response of the detector to changes in the room temperature if the volume of the
detector is constant.

We next consider the small effects that may arise from a volume change due to expansion or 
contraction of the steel walls of the PTB detector as the room temperature increases or 
decreases. The outer case and the bulkheads of the detector chamber are manufactured using 
Mild Steel (Trevor Alloway, Centronic Ltd., private communication, 2009), and typical 
coefficients of linear expansion, $\alpha$, for steel range from $\sim9\times10^{-6}\rm{K}^{-1}$ 
to $17\times10^{-6}  \rm{K}^{-1}$.  
For our analysis here, we will use an average value of 
$13\times10^{-6}  \rm{K}^{-1}$, and assume that the expansion of the 
metal is isotropic.  Based on the stated dimensions of the detector, radius 9.25 cm and 
length 42.7 cm, and neglecting the small volume of the re-entrant tube (well), the 
initial volume, $V_{o}$, of the chamber at the initial temperature can be calculated 
to be $\sim$11477.9 $\rm{cm^{3}}$.  The change in volume, $\Delta V$, arising from a 
change, $\Delta T$, in temperature is
\begin{eqnarray}
\nonumber V_{o}+\Delta{}V &=& \pi\left[ r^{2}\left( 1+\alpha \Delta T\right) ^{2}\right] \left[ L\left( 1+\alpha \Delta T\right) \right] \\
&\cong& V_{o}\left( 1+3\alpha \Delta T\right),
\label{Volume}
\end{eqnarray}
\noindent and hence the new volume of the cylinder will be 11478.3 $\rm{cm^{3}}$ for a 1$^{\circ}$C increase 
in temperature in the room.  The change in volume is thus $\sim$4 parts in 100,000, and would then 
result in a fractional density change of $-4\times10^{-5}$, which is two orders of magnitude smaller than 
the density change proposed in Ref. \citep{Semkow09} to explain the PTB data. Additionally, it should
be noted that this is a 
\textit{decrease} in density, not an increase as claimed in Ref. \citep{Semkow09}, since the same number of gas atoms 
now occupy a negligibly larger volume.  We can therefore conclude that temperature increases 
would lead to lower density of the working medium in the detector, and thus to lower counts in 
the summer, as observed in the PTB data.  However, in order to achieve a $1\times10^{-3}$ fractional change 
in density, which would cause a corresponding fractional change in the ionization and subsequent current, 
the room temperature would need to change by $ \sim $100$^{\circ}$C, which is unrealistic.

In contrast to the operation of the Centronic IG12 ionization chamber employed 
in Ref. \citep{Siegert98} and described above, whose density dependence is 
characterized $\rho \left( T \right) \approx \rm{constant}$  we next consider a hypothetical detector 
in which the pressure, $P$, and the volume, $V$, are held constant, which is the setup 
studied in Ref. \citep{Semkow09}.  In this case, we find from the ideal gas law 
that $\rho T \approx \rm{constant}$, so that $\Delta \rho = -\Delta T/T$.  The change in 
intensity $\Delta I$ of $  \gamma$-rays passing through a thickness $ \Delta x$ of absorbing 
material with density $  \rho$ is given by   $\Delta I/I=\mu\rho\Delta x$ where $\mu$ is the 
absorption coefficient, and this leads to the familiar result \citep{Knoll00} 
$I=I_{o} \rm{exp} \left( -\mu \rho x\right)$. Without loss of generality, we can express 
the counting rate $\left(C\right)$ as  $C=C_{o}\left(1-I/I_{o}\right)$, where $C_{o}$ 
is an appropriate normalization constant, and hence,  
$\Delta C/ \Delta \rho=\mu x \rm{exp}\left(-\mu\rho x\right)$.  To obtain the temperature 
dependence of $C$, we then replace $\left| \Delta \rho \right|$  
by $\left|\rho\right| \Delta T/T$  which gives 
\begin{eqnarray}
\left| \dfrac{\Delta{}C}{C} \right| = \dfrac{\mu\rho x e^{-\mu\rho x}}{1-e^{-\mu\rho x}}\left| \dfrac{\Delta T}{T}\right|.
\label{dendep}
\end{eqnarray}
\noindent Although this reproduces Eq. 1 of Ref. \citep{Semkow09}, we see from the preceding 
discussion that this result only applies to a detector for which  $\rho T = \rm{constant}$. 
This is not the case for the PTB detector, which is characterized by $\rho \left( T \right) \approx \rm{constant}$ 
as noted above. One can illustrate the practical difference between these two cases by using the 
values for $\mu$, $  \rho$, and $x$ quoted in Ref. \citep{Semkow09}:   
$\mu=0.07703\rm{cm^{2}}/g$, $\rho=0.03567 \rm{g/cm^{3}}$,  and $x=14.4 \rm{cm}$.  We find $\mu\rho x =0.0396$, 
and hence, $\left| \Delta C/C \right| = 0.98 \left| \Delta T/T \right|$ .  By way of contrast, 
for the actual PTB detector, the only dependence of $\rho \left( T \right) $  on $T$ is through the small thermal 
expansion of the detector chamber, as discussed above.  Using a coefficient of volume expansion 
$\gamma=39\times10^{-6}(^{\circ}C)^{-1}$ for steel, we find from the preceding 
analysis $\left| \Delta \rho/\rho \right| = 0.98 \left| -\Delta V/V \right|=4\times 10^{-5}$.  
For $T$=300K, we then find for the assumed temperature change in 
Ref. \citep{Semkow09}, $\Delta T=0.91^{\circ}C, \left| \Delta C/C \right| = 3.2\times 10^{-5}$ 
for the actual PTB detector.  This is $\sim$100 times smaller than the 
value  $\left| \Delta C/C \right| = 2.9\times 10^{-3}$ claimed in 
Ref. \citep{Semkow09} from an inappropriate application of the ideal 
gas law to the actual PTB detector.

It follows from the preceding discussion that the expected effects of temperature on the IG12 
ionization chamber used in the PTB measurements are much smaller than the measured 
fluctuations in the $^{226}$Ra data.  We then turn to other possible environmental effects, 
such as humidity, ambient air pressure, or ambient air density, which are less likely to 
affect the detector chamber itself since the detector is a closed system. Moreover, 
these effects are not likely to have an effect on the very small air gap between the 
source and wall of the detector well, since the 4$ \pi \gamma$ ionization chamber is 
measuring the photons (gammas and bremsstrahlung) emitted by the sources, the fluxes 
of which would have little dependence on the marginal air density changes in the small air gap.

\begin{figure}[h]
 \includegraphics[height=68mm] {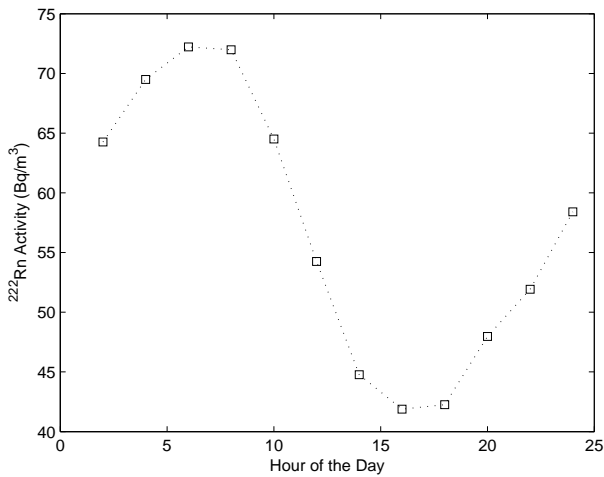}
  \caption{\label{fig:HourlyRn}$^{222}$Rn hourly averaged concentration obtained from Table 1 of Abbady et al. The data were taken at the Centre for Radiation Protection and Radioecology (ZSR) at Hannover University from July 2000 to June 2001.}
 \end{figure}

\begin{figure}[h]
 \includegraphics[height=68mm] {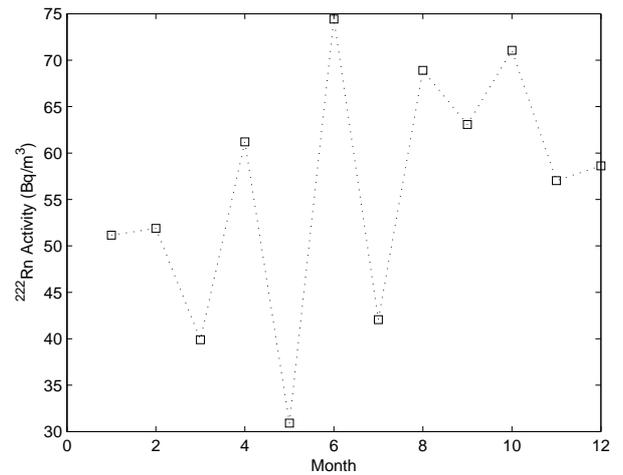}
  \caption{\label{fig:MonthRn}$^{222}$Rn monthly averaged concentration obtained from Table 1 of Abbady et al. The data were taken from June 2000 through May 2001, and are arranged here January through December. See caption to Figure  \ref{fig:HourlyRn} for further details.}
 \end{figure}

We next turn to the possible contributions from background radiation of cosmic and terrestrial 
origins, such as the $^{222}$Rn variations considered by Siegert, et al. \citep{Siegert98}, 
and Schrader \citep{Schrader07} as the cause of the annual variation of 
the $^{226}$Ra measurements. Although $^{222}$Rn concentrations do exhibit seasonal as 
well as diurnal variations \citep{Okabe87,Nishikawa88,Miles88,Gaso94,Thorne03,Abbady04,Wissmann06,Szegvary09}, the actual activities are 
quite small, on the order of 15-90 $\rm{Bq/m^{3}}$, which are far below the activities of the 
sources being measured. Specifically, an analysis performed by Abbady, et al. 
in Hannover, Germany \citep{Abbady04} (which is $\sim$60 km from the PTB in Braunschweig), measured 
indoor concentrations of $^{222}$Rn inside the Centre for Radiation Protection and 
Radioecology (ZSR) at Hannover University. The data were collected for the period of one year, 
with measurements taken every two hours daily, from which monthly averages were calculated 
for each two hour interval. The data, shown in Table 1 of Ref. \citep{Abbady04}, do exhibit a small 
annual variation in the concentration of $^{222}$Rn inside the ZSR offices (an institutional 
building which can be assumed to be somewhat similar to the institutional buildings at PTB), 
with the minimum and maximum concentrations occurring in February and May, respectively. 
Interestingly, the day/night variation shown in Ref. \citep{Abbady04} appears to be 
larger than the annual variation, with the minimum and maximum occurring at 1600 and 0800, 
respectively. Plots of the time-averaged data and the monthly averaged data from Ref. \citep{Abbady04} are shown 
in Figs. \ref{fig:HourlyRn} and \ref{fig:MonthRn}. The diurnal effect is clearly evident, with 
the early morning maximum activity 
concentration possibly happening while the heating, ventilation and air conditioning (HVAC) systems are off or in a lower functional state, as is 
typical in large institutional buildings. It should also be noted that $^{222}$Rn concentrations are known to be dependent on air density, which is in turn dependent on temperature, relative humidity and pressure, as well as on precipitation. It follows that since the warmest part of the day is generally late afternoon, and the coolest is early morning, a diurnal effect is an understandable result. The annual variation is less obvious, but it 
appears that a maximum occurs in Summer/Autumn. 
  
While the preceding information and analyses do not completely rule out an effect caused by $^{222}$Rn, 
it appears that not only are 
the resulting activities quite small, but additionally the phase and period of the $^{222}$Rn 
concentrations do not match those of the $^{226}$Ra measurement variations reported 
in Ref.~\citep{Siegert98} and later analyzed in Ref.~\citep{Jenkins09Capp}.

Other contributions to background in the PTB measurements would be of terrestrial and cosmic 
origins, such as muons, or photons from $^{40}$K decay. Efforts were made by Siegert, et al. to reduce the 
background by use of significant lead shielding around the detector \citep{Siegert98,Schrader00}. 
However, it should be noted that the $^{226}$Ra data provided by Schrader (H. Schrader, private communication, 2006)
for the analysis presented in Ref.~\citep{Jenkins09Capp} were 
already corrected for background, which obviates the need for analyses of these background 
effects. Nevertheless, a discussion of those possible backgrounds is presented below. An 
analysis of environmental radiation at ground level was conducted by Wissmann at the 
PTB \citep{Wissmann06}, utilizing data from the Cosmic Radiation 
Dosimetry Site, situated on a lake near the PTB laboratory. The results of the analysis, which 
corrected for fluctuations in weather and solar effects, exhibited an annual variation 
of $\pm{}6.9 \rm{nSv} \cdot h^{-1}$, which is equivalent to a dose variation equal 
to $\pm{}1.92\times10^{-3} \rm{nSv} \cdot s^{-1}$. Given the initial amount of $^{226}$Ra in the 
PTB source, 300 $\mu{}$g, which would have had an initial activity of 11.1 MBq, the activity would be
$\sim$88.8 MBq after 40 years due to the ingrowth of the radioactive daughters. Utilizing the more conservative
activity value of 11.1 MBq, the gamma dose 
rate from the source can be estimated to be $\sim2.1\times10^{4} \rm{nSv} \cdot s^{-1}$, which is 7 orders 
of magnitude larger than the variation in the cosmic dose rate reported in 
Ref. \citep{Wissmann06}.  We can therefore 
conclude that seasonal variations in cosmic radiation background would likely have little if any effect 
on the PTB detector system.

Returning briefly to the background radiation from $^{222}$Rn, it should also be noted since $^{226}$Ra decays to $^{222}$Rn, the decay chain 
of $^{222}$Rn is identical to the decay chain of $^{226}$Ra.  If the $^{222}$Rn daughters 
are in equilibrium with the relatively constant $^{222}$Rn concentration of $\approx55$ Bq, the 
activity of $^{222}$Rn and its daughters at any time will be $\approx300 \mathrm{Bq}$, which equates to a 
gamma dose rate of $\approx0.58 \mathrm{nSv} \cdot \mathrm{s^{-1}}$. This value is five orders of magnitude smaller than the 
dose rate of the $^{226}$Ra source itself. Hence, it can be safely assumed that even with 
a variation of the atmospheric $^{222}$Rn concentration of $\pm{}100\%$ (which is significantly larger 
than the actual variation noted in Ref. \citep{Abbady04}), the even larger activity of $\approx600\mathrm{Bq}$ would 
yield a new gamma dose rate of $\sim 1.16~\mathrm{nSv}\cdot \mathrm{s^{-1}}$, which is still four 
orders of magnitude smaller than the dose from the $^{226}$Ra source itself. This effectively 
rules out a discharge effect on cables, etc. arising from charged particles or photons 
produced by $^{222}$Rn daughters as the cause of the measured variations reported in Ref. \citep{Siegert98}.

\section{The PTB Solid State Detector}

The $^{154}$Eu and $^{155}$Eu samples studied by Siegert, et al. \citep{Siegert98} 
were measured with three different semi-conductor detectors as part of the evaluation of 
detector system stability previously discussed.  In the course of these measurements, a 
separate $^{152}$Eu calibration source was used to measure the changes in efficiency of 
the Ge(Li) detector which was one of the three semi-conductor detectors examined.  A plot of 
measurements of the efficiency of a Ge(Li) detector versus time, which is presented 
in Fig. 2 of Ref. \citep{Siegert98}, exhibits a similar period of 1 year, shifted by 
half a period (or about 6 months), with an amplitude of about 0.5\%.  The 
efficiency of a detector is determined by comparison of the measured counts taken with the detector 
to the expected output of a known source in a fixed geometry.  Thus, for the Ge(Li) measurement of 
the $^{152}$Eu point source, the minima of the efficiency curve represent a deficit in counts 
(with respect to the expected counts based on the known activity and half-life of the standard), 
while the maxima represent an excess of counts. Specifically, the minima of the efficiency plot fall near the 
beginning of the calendar year for the Ge(Li) detector, which is opposite to the signal seen 
in the ionization chamber measurements of $^{226}$Ra.  This appears to call into question the conclusion of Jenkins, 
et al. \citep{Jenkins09Capp} that decay rates are simply dependent on the distance 
of the Earth from the Sun. However, the fact that the period of the $^{152}$Eu data is the same, but 
shifted by half a period, still suggests a possible link to the Earth-Sun distance.

This possibility is supported by an examination of the decay modes of $^{152}$Eu which has 
two branches, K-capture (72.10\%) and $\beta{}^{-}$ decay (27.90$\%$) \citep{Artna-Cohen96} .  The 1408 keV photon 
measured with the Ge(Li) detector \citep{Siegert98} only arises from the K-capture branch, and is seen in 21.07\% 
of the decays (there is no photon of that energy in the $\beta{}^{-}$ decay branch)\citep{Artna-Cohen96}.  In contrast 
to the Ge(Li) detector, the ionization chamber is incapable of discriminating among the different 
photons measured, and hence could not track the 1408 keV photon separately, as it would be lost 
in the sum of all the photons emitted by the sample. The phase of the $^{152}$Eu data could thus be understood
if K-capture modes responded to an external influence oppositely from $\beta{}^{-}$ decays.

In Refs. \citep{Jenkins09Fapp,Jenkins09Capp} 
the conjecture was put forward that a field or particle emanating from the Sun, possibly neutrinos, might 
be enhancing or interfering with the decay of radioactive nuclides.  
Since the quantum mechanical details of the K-capture and $\beta{}^{-}$ decays are quite different, 
it is in fact possible that the response of the K-capture branch would be different from that 
of the $\beta{}^{-}$ decay branch.  If these were in fact opposite effects, i.e. the decay rate 
for the $\beta{}^{-}$ decay branch would be increasing when the K-capture rate was decreasing, 
these competing effects would tend to damp out any fluctuations. That such a possibility 
could actually occur is supported by data from $^{54}$Mn (which also decays via K-capture) 
acquired during the solar flare of 2006 December 13, \citep{Jenkins09Fapp}. It was 
observed that the $^{54}$Mn count decreased from the expected rate during the three-day period encompassing the 
solar flare. Thus, if we assume that the solar neutrino flux 
increased during the solar flare, then the response of $^{54}$Mn and $^{152}$Eu to increased 
solar flux would have the same phase in the sense that both decay rates decreased in response to an increase in 
solar neutrino flux.

\section{The BNL Proportional Counter}

We next turn to an examination of the environmental sensitivity of the gas proportional 
counter and sample changing system used by Alburger, et al. \citep{Alburger86} and 
Harbottle, et al. \citep{Harbottle73} to measure the half-life of $^{32}$Si. The 
detector in question was  cylindrical, $\sim$1.5 in. diameter, 2 in. depth, with a 1 in. 
end window, operating on P-10 gas. The detector pressure was held constant by a device on the 
gas outlet vented to a barostated enclosure to limit fluctuations in the density of the P-10 gas. 
The detector itself was 
mounted on an automatic precision sample changer, which was entirely contained within the 
aforementioned barostated enclosure, as described in Refs. \citep{Alburger86,Harbottle73}. 
The window material was 0.006mm Kapton$^{\circledR}$ (Dupont H-film), with 
gold vacuum-deposited on both sides (40-50~\AA{} on the outer surface, 120-150~\AA{} on the inner 
surface). A bias voltage of 2150 V was applied to the detector, which was the voltage closest to the center 
of the beta counting plateau \citep{Alburger86}.

Gas proportional detectors utilize gas multiplication, where free electrons generated 
by the ionization of the gas by the incoming particles are accelerated by the strong 
electric field (created by the bias voltage) to sufficient kinetic energy to cause further 
ionizations. Electrons freed in these ionizations can also be accelerated to cause 
additional ionizations, as long as their energy is greater than the ionization energy of 
the neutral gas molecules. This gas multiplication process, or cascade, is known as a Townsend 
avalanche, which terminates when all free electrons are collected at the 
anode \citep{Knoll00}. With a well designed and maintained detector, the 
number of secondary ionizations is proportional to the number of initial ionizations, but 
multiplied by a factor of ``many thousands''\citep{Knoll00}. Almost all primary ion pairs are formed outside 
the multiplying region, which is confined to a small volume around the anode, and the 
electrons from the primary ionizations drift to that region before multiplication takes 
place, ``therefore, each electron undergoes the same multiplication process regardless of 
its position of formation, and the multiplication factor will be [the] same for all original 
ion pairs.'' \citep{Knoll00} If the charge pulse of all of the electrons 
collected at the anode per event is larger than the discriminator level setting after 
amplification, a count is then recorded by the scaler.

The two sources used in the BNL experiment, $^{32}$Si and $^{36}$Cl, are described in detail in 
Alburger, et al., Sec. 2 of Ref. \citep{Alburger86}. The isotope $^{32}$Si 
undergoes $\beta^{-}$-decay (100$\%$, $E_{\beta{}max} =224.31\rm{keV}$) to $^{32}$P 
with a T$_{1/2}$=157(19) y \citep{Singh06}. The radioactive daughter, $^{32}$P, (which is 
in secular equilibrium with $^{32}$Si) undergoes $\beta{}^{-}$-decay (100$\%$, 
$E_{\beta{}max}$ =1710.5(21) keV) to $^{32}$S (stable) with a $T_{1/2}$=14.262(14) d \citep{Endt99P32}. The isotope $^{36}$Cl, on the other hand, decays 
via competing $\beta{}$-decay branches ($\beta{}^{-}$, 98.1(1)$\%$, $E_{\beta{}max}$=708.6(3) 
keV; $\beta{}^{+}/\epsilon, 1.9(1)\%)$ \citep{Endt99Cl36}. Great care was 
taken by the BNL group in preparation of the $^{32}$Si source to ensure that no 
contaminants would be included that would confuse the data collection and analysis. 
Analyses of the spectra of each of the sources were conducted with a plastic scintillator 
to test for impurities or other problems, and the spectra were as expected. The $^{32}$Si 
betas were not noticeably evident in the spectrum from the plastic scintillator, which 
was likely due to the effective thickness of the sample (17 mg/$\rm{cm^{3}}$) that served to 
degrade and smear the low-energy betas, as noted by Alburger, et al. \citep{Alburger86}. 
However, they did appear with the expected equal distributions when part of the same 
material was measured in a liquid scintillator \citep{Alburger86,Cumming83}. The $^{36}$Cl 
spectrum was as expected, as reported by Alburger, et al. \citep{Alburger86}.

The gas proportional detector described above was utilized in a differential counter system 
described in detail in Ref. \citep{Harbottle73}, which used a 
precision sample changing system that would alternate the sources during counting runs. 
The system was designed and built to allow maximum reproducibility, e.g. by utilizing 
precision micrometer heads which adjusted the source/detector distance to 
within $\pm{}$0.001mm \citep{Harbottle73}. As stated before, this system was contained 
within a ``pressure-regulated box in order to minimize the differential energy-loss effects 
that would occur with changes in barometric pressure'' \citep{Alburger86}. 
Tests of the sensitivity of the counting system to changes in the box pressure, as well 
as to detector bias voltage, detector gas flow, and discriminator setting were carried out 
prior to the series of measurements, and the estimated effects on the $^{32}$Si counts 
and the Si/Cl ratio are detailed in Table 1 of Ref. \citep{Alburger86}. 

\begin{figure}[h]
 \includegraphics[height=67mm] {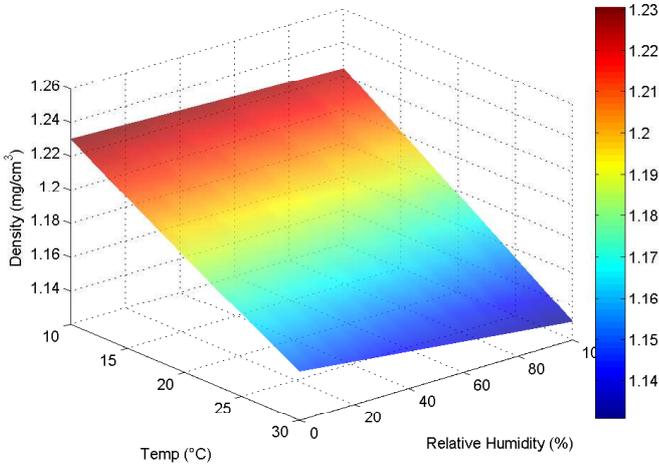}
  \caption{\label{fig:air-den}Plot of air density as a function of temperature and relative humidity. The data were calculated based on the formulae presented in Refs. \citep{Giacomo82,Davis92,Picard08}}
 \end{figure}

Notwithstanding the tests described above, Alburger, et al. identified an unexpected periodic 
fluctuation in the Si/Cl ratio with a period of approximately 1 year. Having virtually eliminated 
other systematic variables, or at least having controlled them to the extent possible, 
Alburger's BNL group was left with temperature and humidity as possible explanations, which 
are known to change the density of air (even if the pressure of the box containing the 
detector system is held constant). Increases in temperature will decrease the density 
of air, as will increases in relative humidity, although humidity has a much smaller effect 
than temperature. To summarize, warmer more humid air, i.e. summer air, is less dense than 
cooler, drier winter air. This is depicted in Fig. \ref{fig:air-den}, which plots the temperature and 
relative humidity dependence of the density of air, based on equations developed and 
recommended by the Comit\'{e} International des Poids et Mesures (CIPM) \citep{Giacomo82,Davis92,Picard08}.
 
The Alburger group at BNL did not begin collecting data on temperature and humidity until 
the last five months of the four year experiment, and during that time saw a range of 
temperatures of 72.4-74.7$^{\circ}$F, and a relative humidity range of 35-76\% \citep{Alburger86}. 
However, further investigations into data collected in other laboratories within the same 
building suggested that the average temperatures remained well within the range of 70-76$^{\circ}$F 
year round. The BNL group noted that the fluctuations appear to follow the same annual 
cycle as outdoor temperature. They also noted that if the indoor temperature and relative humidity 
were to track the outdoor conditions, then the data appeared to indicate that the fluctuations 
were environmentally based. The cooler, drier, and denser winter air could attenuate the lower 
energy $^{36}$Cl betas, and would thus affect the ratio Si/Cl because the lower energy $^{36}$Cl 
counts would be more weather dependent. However, Alburger, el al. concluded that ``in order to 
produce the variations of $\pm{}$3 standard deviations, the large humidity changes would have to 
be combined with temperature variations over a range of at least $\pm{}5^{\circ}$F, which is 
larger than the probable actual range. We therefore conclude that systematic periodic variations 
are present but that they cannot be fully accounted for by our tests or 
estimates''\citep{Alburger86}.

\begin{figure}[h]
 \includegraphics[height=67mm] {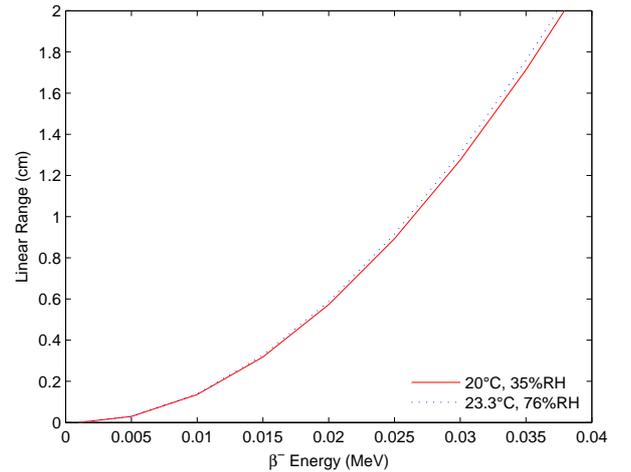}
  \caption{\label{fig:ESD}Linear range of betas in air as a function of beta energy. The graphs are calculated using the results of Katz and Penfold in Ref. \citep{Katz52}. The air densities shown are based on the maximum and minimum values of temperature and humidity quoted by Alburger, el al. in Ref. \citep{Alburger86} }
\end{figure}

The conclusions of the BNL group can be supported by estimating the effect of temperature and relative humidity on the range of betas over 
the spectrum of energies that would be seen from both sources. The range of an electron 
can be estimated by utilizing an analysis by Katz and Penfold \citep{Katz52}, 
which provided a range-energy $\left( R-E_{o} \right)$ relationship given by:
\begin{eqnarray}
R\left( \rm{mg/cm^{3}} \right) =412 E_{o}^{1.256-0.0954 \ln E_{o}}
\label{electron}
\end{eqnarray}
\noindent The linear range can be derived by dividing $R$ by the density of the propagating 
medium, air, and expressing the electron energy, $E_{0}$ in MeV (it should be noted that these linear ranges
are not necessarily straight line paths).  We utilize the variations suggested 
by Alburger, et al. for 
temperature (T) (73$\pm{}$5$^{\circ}$F) and relative humidity (RH) (35-76\%), and assume 
that the maximum density would be associated with the minimum T and RH (68$^{\circ}$F, RH=35\%), 
and the minimum air density with the maximum T and RH (78$^{\circ}$F, RH=76\%). The air 
densities are then found to be 1.185mg/$\rm{cm^{3}}$ and 1.155 mg/$\rm{cm^{3}}$, respectively. The energy-dependent
linear ranges of energetic electrons for each condition set are shown in Fig. \ref{fig:ESD},
which gives the linear range of betas of various low energies in air. We have neglected the effects of 
the energy losses due to transport through the detector window or sources, which are assumed to 
be independent of temperature. It should be noted again, that the $^{36}$Cl source was 4mm from 
the detector window, and the $^{32}$Si source 1mm away. Fig. \ref{fig:ESD} confirms the conclusions of the BNL group that 
changes in T and/or RH would have been too small to explain the fluctuations in their data. These conclusions are
further supported by an MCNP analysis which we have carried out, the results of which will be described below.
Before turning to this analysis, we first consider the remaining question of backgrounds which could impact the BNL detector system. 

The BNL group measured an initial background counting rate of 6.6(3)/min. 
This background measurement was repeated after the conclusion of all runs by counting with 
the $^{32}$Si source removed from its holder, and the $^{36}$Cl still in place (to ensure that 
there was no cross-talk between the sources). The rate was found to be 6.5(3)/min. Since the average 
count rates for the measurements were initially 
substantially larger, 21,500/min and 14,800/min for the $^{36}$Cl source and $^{32}$Si source, 
respectively, the background count rates were considered insignificant, and hence were ignored by the BNL group. 
Additionally, since the sources were alternated during the counting procedure for 
each run, 30 minutes each for 20 cycles, the background would likely be the same for both of the 
isotopes.

\section{Monte Carlo Modeling of the BNL Detector}

The Monte Carlo radiation transport software package, MCNPX \citep{Pelowitz08}, was used to determine the effects 
of changes in ambient air temperature on the proportional detector system used by Alburger et al. \citep{Alburger86}.  The air density 
between the source and detector window was varied over the temperature range 40$^{\circ}$F-90$^{\circ}$F by use 
of the ideal gas law, corresponding to an approximate density range between
1.1350$\times10^{-3} \rm{g/cm^{3}}$ and 1.2486$\times10^{-3} \rm{g/cm^{3}}$ for dry air. While humidity 
also affects total air 
density, this effect is presumed to be small (see Fig. \ref{fig:air-den} ) relative to the effects of temperature 
change, and hence was ignored in this study. Additionally, the 40-90$^{\circ}$F range that was analyzed 
encompasses both the minimum and maximum air densities expected for a wide range of both 
temperature and humidity in the laboratory. The physical parameters of the detector 
system (including the proportional detector, $^{32}$Si source, and $^{36}$Cl source) were 
obtained from the literature \citep{Alburger86,Harbottle73}, the BNL group's experimental 
notes (private communication, D. Alburger and G. Harbottle, 2005), and measurements of both the actual 
sources used in the experiment and a proportional detector of similar design.

The cylindrical end-window proportional detector was modeled as a stainless steel (type 304) 
cylindrical shell with inner diameter 3.96875 cm, length 5.08 cm, and 3 mm wall thickness. The 
detector window was modeled following  Harbottle et al. \citep{Harbottle73} 
as 0.006 mm Kapton$^{\circledR}$ plated with vacuum-deposited gold on both the inner (125 $\AA$) and 
outer (45 $\AA$) surfaces. The detector volume was filled with P-10 gas (90\% argon, 10\% methane 
by volume) with a calculated density of 1.6773$\times10^{-3} \rm{g/cm^{3}}$.  This gas density within the 
detector was held constant.

Due to the possibility that the lower energy betas from $^{36}$Cl $(708.6 \mathrm{keV})$ were more easily affected by changes
in air density than the higher energy $^{32}$Si-$^{32}$P betas $(1710.5\mathrm{keV})$, each source was modeled separately, which also allowed us to account for the difference in proximity of each
source to the detector.  The $^{32}$Si 
source, which was modeled in detail as described by Alburger et al. \citep{Alburger86}, consisted 
of a brass base 3.18 cm in diameter and approximately 0.3175 
cm (1/8 in.) thick with a 1.91-cm-diameter, 0.8-mm-deep recess. The recess contained 47.7 mg of 
SiO$_2$ covered by 9.263$\times10^{-5}$ cm of aluminum foil. The MCNPX electron source volume was defined 
uniformly over the volume occupied by the SiO$_2$. According to Alburger et al., the $^{32}$SiO$_{2}$ 
source was created approximately 13 years prior to the start of the half-life 
experiment \citep{Alburger86}, and hence $^{32}$Si was certainly in secular equilibrium 
with its radioactive daughter, $^{32}$P, during the experimental time frame. The source energy spectrum 
used in these MCNPX models consisted of equal contributions from the $^{32}$Si (E$_{max}$ = 225 keV) 
and $^{32}$P (E$_{max}$ = 1710.5 keV) beta energy spectra calculated according to Fermi theory, which
allowed us to incorporate the loss of the much lower energy betas from $^{32}$Si.

As was the case for the $^{32}$Si source, the $^{36}$Cl source was placed on a brass base 3.18 cm 
in diameter and 0.3175 cm thick, but without a recess machined into the top. The $^{36}$Cl was 
spread over a diameter of 1.91 cm in the center of the top surface of the base. Thus, the MCNPX electron 
source was defined as a flat disk 1.91-cm in diameter on the surface of the brass base with an 
energy spectrum equal to that of $^{36}$Cl (E$_{max}$ = 709.2 keV) calculated according to Fermi theory.

According to the BNL experimental notes, the distance of each source from the detector window 
was adjusted in order to obtain similar count rates from each, and this was achieved by having the $^{32}$Si source 
1.000 mm from the detector window, while the $^{36}$Cl was placed at a distance of 4.000 mm.  
The source position in the MCNPX models for each source reflected these distances.

As stated before, Alburger, et al. reported \citep{Alburger86} that the temperature variation was 
well within the range 70-76$^{\circ}$F. We found that tally values varied so little over this 
range that they were statistically constant to within the minimum achievable relative error 
in MCNPX. Thus, a large temperature range of 40$^{\circ}$F to 90$^{\circ}$F was chosen to improve the statistical 
determination of the function relating each tally to air temperature (and associated density). 
The results presented here were obtained using $1\times10^{8}$ (for $^{36}$Cl) 
or $2\times10^{8}$ (for $^{32}$Si-$^{32}$P) source electrons for each density variation, and this 
provided relative errors of $\sim10^{-4}$ for the particle/energy current tallies, and $\sim2\times10^{-4}$ 
for the energy deposition tallies.

A number of MCNPX tallies were used to test the assertion that variations in air density with 
temperature would result in an associated increase or decrease in the probability that an electron 
with a given energy will traverse the air gap between the source and detector window. A 
linear least-squares fit was applied to the results of each tally as a function of air 
temperature, and these results are summarized in Table 1.

\begin{table*}

\label{table1}
\caption{Table of MCNPX results showing the sensitivities of the system per degree Fahrenheit. }

\begin{tabular}{|p{3.5cm}|l|c|c|c|c|}
\hline & \multicolumn{2}{|c|}{$^{32}$Si-$^{32}$P} & \multicolumn{2}{|c|}{$^{36}$Cl} &  ($^{32}$Si-$^{32}$P)/$^{36}$Cl \\
  & Per Source $e^-$  & Norm.(70$^{\circ}$F)& Per Source $e^-$& Norm.(70$^{\circ}$F)& Ratio \\
\hline  Det. E Deposition (MeV/ptcl) ($\Delta/^{\circ}$F) & $0.051(24)\times10^{-6}$ & $10.1(67)\times10^{-6}$ & $0.34(40)\times10^{-6}$ & $39.8(67)\times10^{-6}$ & $-17.7(40)\times10^{-6}$  \\ 
\hline Det. Window $e^-$ Current ($\Delta/^{\circ}$F) & $4.39(86)\times10^{-6}$ & $12.1(33)\times10^{-6}$ & $20.3(13)\times10^{-6}$ & $35.3(33)\times10^{-6}$ & $-14.6(21)\times10^{-6}$ \\ 
\hline Det. Window E Current(MeV/ptcl) ($\Delta/^{\circ}$F) & $1.48(45)\times10^{-6}$ & $7.8(33)\times10^{-6}$ & $4.32(32)\times10^{-6}$ & $32.4(34)\times10^{-6}$ & $-35.2(48)\times10^{-6}$ \\ 
\hline Det./Source $e^-$ Current Ratio ($\Delta/^{\circ}$F) & $6.8(32)\times10^{-6}$ & $7.1(48)\times10^{-6}$ & $29.5(24)\times10^{-6}$ & $41.5(48)\times10^{-6}$ & $-46.5(64)\times10^{-6}$ \\ 
\hline Det./Source E Current Ratio ($\Delta/^{\circ}$F) & $5.1(33)\times10^{-6}$ & $5.2(47)\times10^{-6}$ & $24.9(25)\times10^{-6}$ & $33.7(48)\times10^{-6}$ & $-37.9(63)\times10^{-6}$ \\ 
\hline 
\end{tabular} 
\end{table*}

\begin{figure}[h]
 \includegraphics[height=70mm] {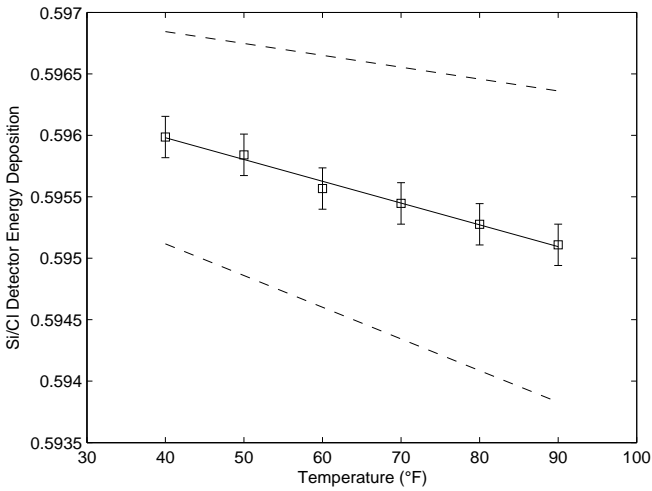}

 \caption{\label{fig:SiClRatEDep}The ratio of $^{32}$Si to $^{36}$Cl energy deposition in the proportional detector. Dashed lines indicate the 95\% confidence interval for the calculated linear least-squares fit of the data.}
 \end{figure}

\begin{figure}[h]
 \includegraphics[height=70mm] {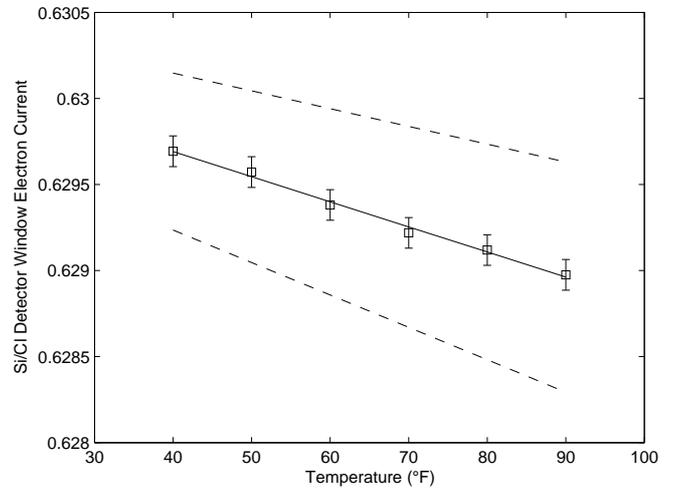}

 \caption{\label{fig:RatWinelecCurr}The ratio of $^{32}$Si to $^{36}$Cl electron current (number of electrons) across the proportional detector window (into the detector). Dashed lines indicate the 95\% confidence interval for the calculated linear least-squares fit of the data.}
 \end{figure}

\begin{figure}[h]
 \includegraphics[height=70mm] {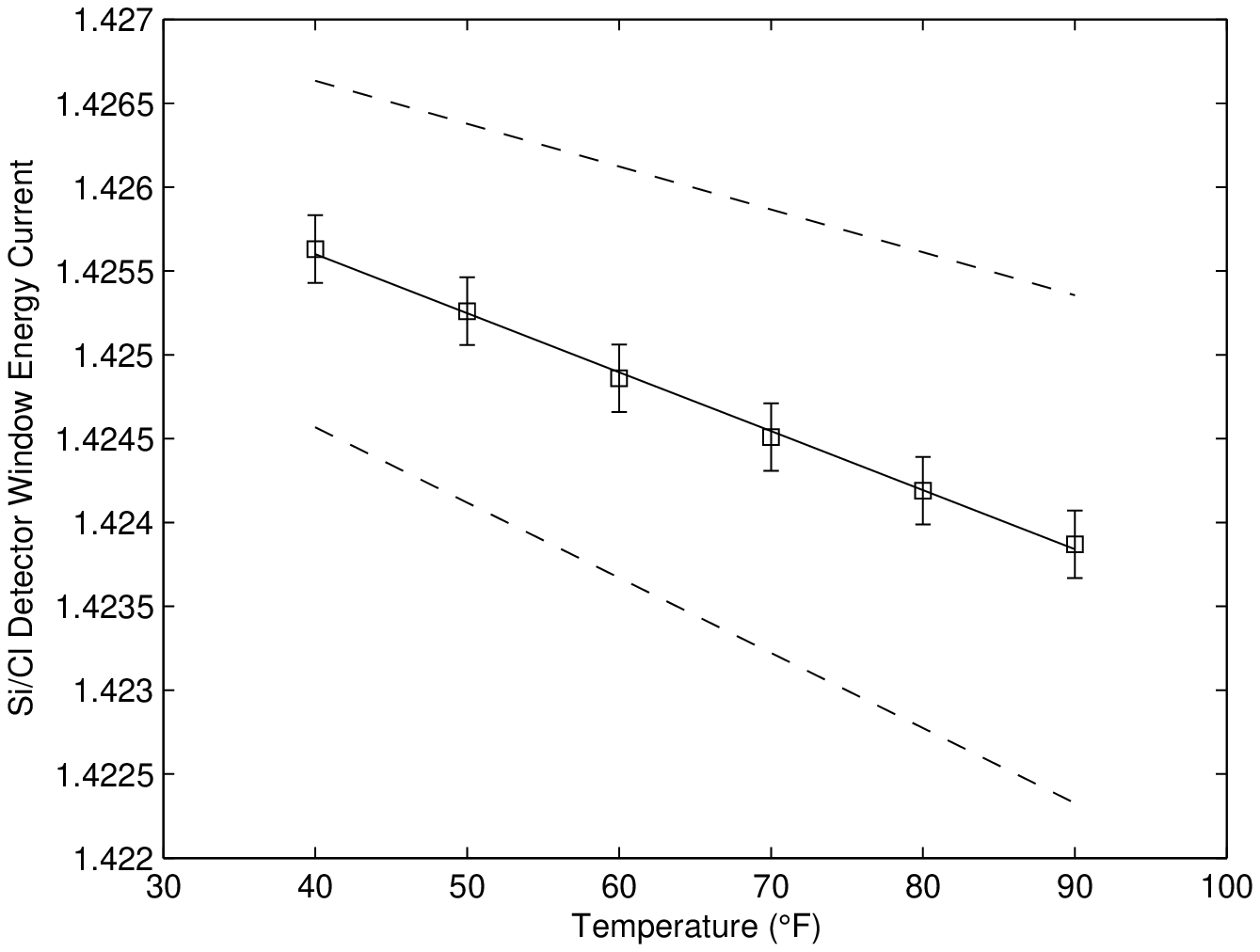}

 \caption{\label{fig:RatWinEnCurr}The ratio of $^{32}$Si to $^{36}$Cl energy current (MeV) across the proportional detector window (into the detector). Dashed lines indicate the 95\% confidence interval for the calculated linear least-squares fit of the data.}
 \end{figure}

%
Of particular interest are the temperature dependent variations in the energy deposition within
the detector (Fig. \ref{fig:SiClRatEDep}) , the electron 
current at the detector window (Fig. \ref{fig:RatWinelecCurr}) , and the energy current at the detector
window (Fig. \ref{fig:RatWinEnCurr}) , all of which can be used as 
surrogates for detector count rate. For each of these tallies, ratios of the value per source 
electron for $^{32}$Si to that of $^{36}$Cl were 
calculated.  Energy deposition, 
electron current, and energy current ratios were found to vary 
by $1.77(40) \times 10^{-5}/^{\circ}$F, $1.46(21)\times10^{-5}/^{\circ}$F, 
and $3.52(48)\times10^{-5}/^{\circ}$F, respectively. These results are echoed by 
the tally variations as a function of temperature for $^{32}$Si-$^{32}$P and $^{36}$Cl individually 
in Table 1. As expected, all individual tally values increase with increasing temperature due 
to the associated decrease in air density between the source and detector. In all cases, the rate 
of change is greater for $^{36}$Cl due to its lower energy spectrum (relative to $^{32}$P) and greater source-detector
distance. In general, 
detector events arising from the $^{32}$Si source are due almost entirely to $^{32}$P betas because 
the low-energy $^{32}$Si betas are largely attenuated by the source itself \citep{Alburger86}. This explains the negative slope associated with the Si/Cl ratios as temperature increases.

\begin{figure}[h]
 \includegraphics[height=70mm] {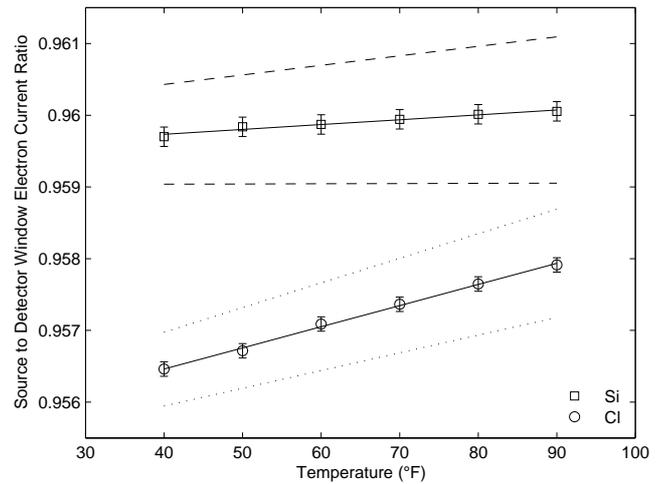}
  \caption{\label{fig:SourceDetWinelecCurr}The ratios of electron current exiting the source toward the detector and electron current across 
 the proportional detector window (into the detector) for $^{32}$Si and $^{36}$Cl. $^{36}$Cl data 
 have been shifted up by 0.2465 to facilitate visualization of both curves. Dashed and dotted lines 
 indicate the 95$\%$ confidence interval for the calculated linear least-squares fit of the data 
 for $^{32}$Si and $^{36}$Cl, respectively.}
 \end{figure}

\begin{figure}[h]
 \includegraphics[height=70mm] {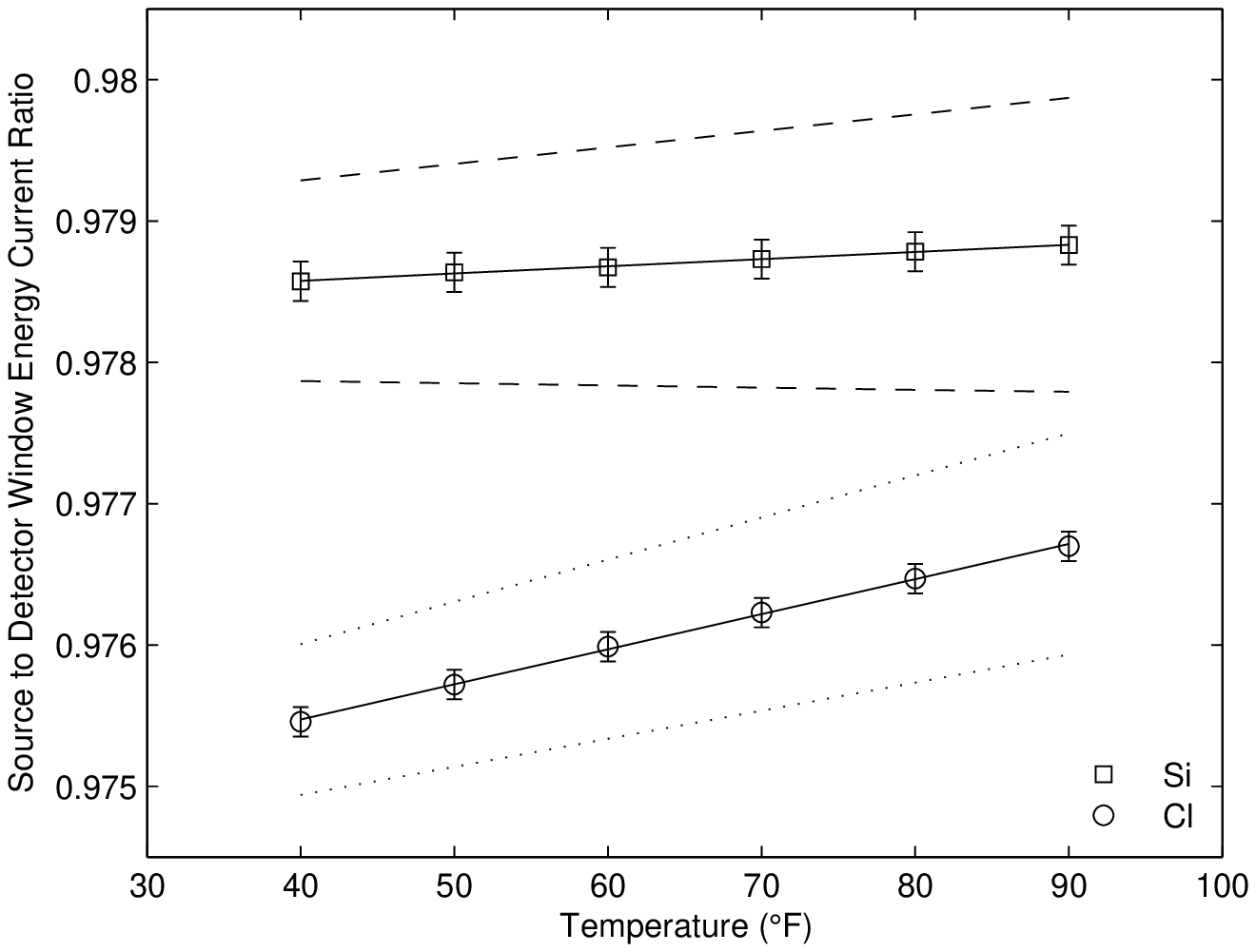}
  \caption{\label{fig:SourceDetWinEnergyCurr}The ratios of energy current exiting the source toward the detector and energy current across 
 the proportional detector window (into the detector) for $^{32}$Si and $^{36}$Cl. The $^{36}$Cl data 
 have been shifted up by 0.2390 to facilitate visualization of both curves. Dashed and dotted lines 
 indicate the 95\% confidence interval for the calculated linear least-squares fit of the data 
 for $^{32}$Si and $^{36}$Cl, respectively.}
 \end{figure}

The ratios of electron current and energy current across a plane just above the source, and above the 
detector window were also examined (see Figures \ref{fig:SourceDetWinelecCurr} and \ref{fig:SourceDetWinEnergyCurr} ). These ratios directly address the 
assertion that changes in air density result in count rate variations due to increased or decreased 
absorption of electrons in the air gap between source and detector. The $^{32}$Si electron and energy 
current ratio dependence was found to be $6.8(32)\times10^{-6}/^{\circ}$F 
and $5.1(33)\times10^{-6}/^{\circ}$F, respectively, and for $^{36}$Cl, 
$2.95(24)\times10^{-5}/^{\circ}$F and $2.49(25)\times10^{-5}/^{\circ}$F, 
respectively. These temperature dependencies are all approximately 2 orders of magnitude smaller 
than the periodic variations observed by Alburger et al.

As we noted previously, following a number of experimental tests, Alburger et al. concluded that it was possible that part 
of the 0.3\% annual variation 
in observed count rates could be attributed to changes in ambient temperature (between 70 and 76 $^{\circ}$F), and 
relative humidity (35 to 76\%) \citep{Alburger86}. However, the 
results obtained in this Monte Carlo analysis indicate that a temperature variation much larger 
than 6$^{\circ}$F would be required to generate the observed count rate oscillations. Given 
that the ratio of energy deposition in the detector varies by approximately $1.2\times10^{-5}/^{\circ}$F, 
the temperature of the air separating the source from the detector would have had to vary by well over 
100 $^{\circ}$F to account for the observed data. We therefore conclude that changes in temperature 
and relative humidity alone could not have caused the observed periodic count rate variations.
\section{Further Analysis}

\begin{figure}[h]
 \includegraphics[height=67mm] {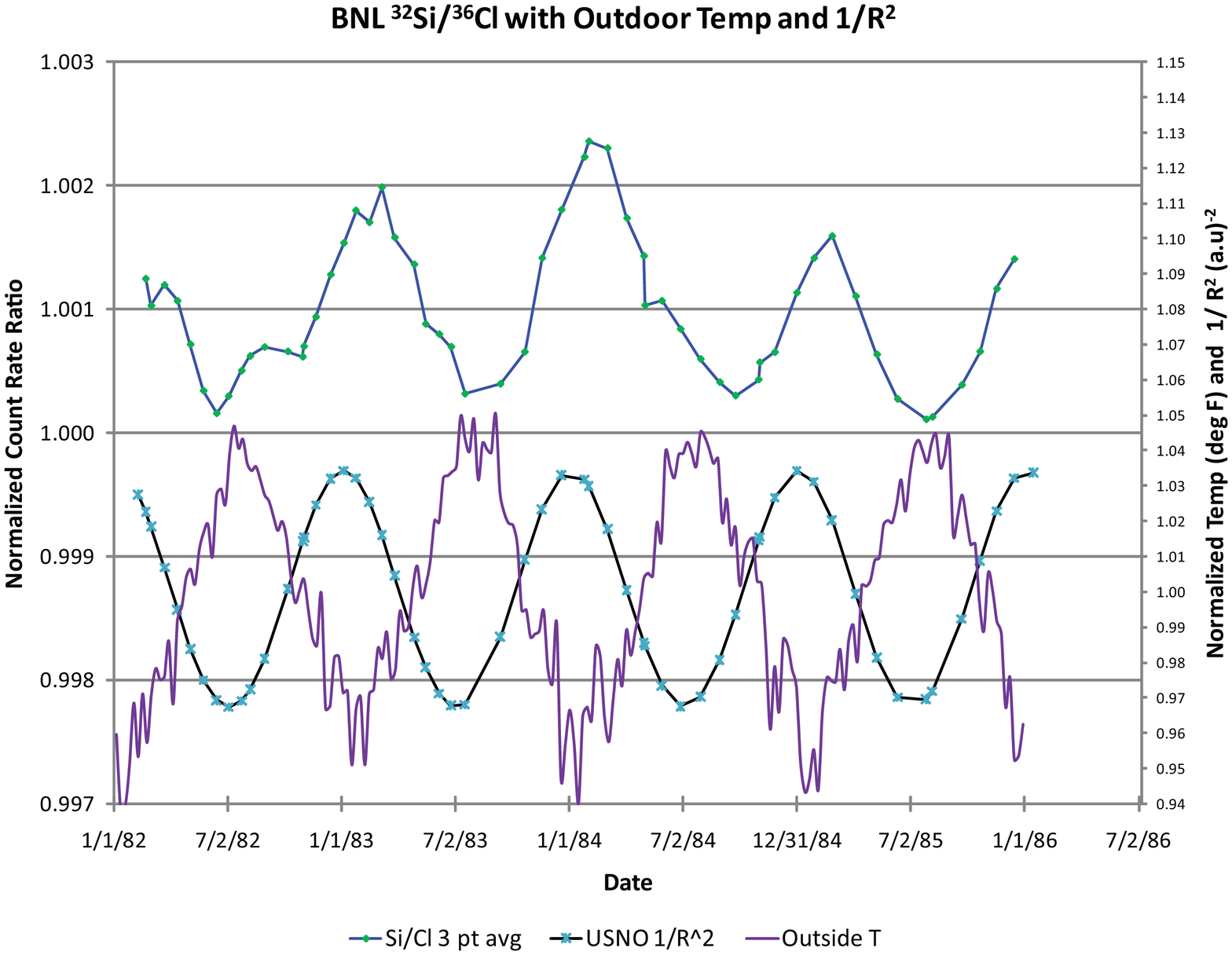}
  \caption{\label{fig:SiClratio}The counting rate ratio $ \dot{N}\left(\rm{Si}\right)/\dot{N}\left(\rm{Cl}\right) $ for the BNL data as a function of time, along with $ 1/R^2 $ and the outdoor temperature measured by NOAA near BNL. The correlation with $ 1/R^2 $ and the anti-correlation with temperature are both evident. See text for further discussion.}
\end{figure}

In Fig. \ref{fig:SiClratio} , we exhibit the count 
rate ratio $\mathrm{Si/Cl}\equiv\dot{N}\left(\rm{Si}\right)/\dot{N}\left(\rm{Cl}\right)$ as a 
function of time, along with the outdoor temperature variation for the area surrounding the BNL \citep{NCDC}
and $1/R^2$.  We see that Si/Cl approximately correlates with $1/R^2$ and anti-correlates with 
temperature, in both cases with observable phase shifts.  Since at this stage we have no theoretical 
reason to suppose that Si/Cl should correlate (rather than anti-correlate) with $1/R^2$, both 
of the correlations in Fig.~\ref{fig:SiClratio} are significant.  However, we have already ruled out temperature 
as the primary explanation of the Si and Cl individual fluctuations, and hence we wish to 
understand whether influences emanating from the Sun, such as neutrinos, could be compatible with these data.

In such a picture, the effect of analyzing the ratio Si/Cl is to largely cancel out small fluctuations in 
voltage, currents, temperature, background, etc., which are common to 
both $\dot{N}\left(\rm{Si}\right)$ and $\dot{N}\left(\rm{Cl}\right)$, 
while leaving intact the presumably larger effects due to the variation of $1/R^2$.  From this 
perspective, the fact that the Si/Cl ratio does not exactly track $1/R^2$ is understandable in terms of the differing 
sensitivities of Si and Cl to external perturbations:  the same nuclear structure mechanisms 
which endow Si and Cl with very different lifetimes (172y and 301,000y, respectively) could 
presumably result in different responses of these nuclei to small perturbations. Evidence of this effect
was first pointed out by Ellis \citep{Ellis90}, who saw an annual periodic signal in $^{56}$Mn,
but not in $^{137}$Cs when both nuclides were measured for several years on the same detector system. 
Therefore, a long-term periodic signal due to the variation of $1/R^2$ would survive in the ratio of Si/Cl.  
This can be understood quantitatively by writing 
\begin{eqnarray}
\nonumber \dfrac{\dot{N}\left(\rm{Si}\right)}{\dot{N}\left(\rm{Cl}\right)} 
&=&\dfrac{A\left( \rm{Si} \right)\left[1+\epsilon\left(\rm{Si}\right)\cos{\omega t}\right]}{A\left( \rm{Cl} \right)\left[1+\epsilon\left(\rm{Cl}\right)\cos{\omega t}\right] } \\
&\cong& \dfrac{A\left(\rm{Si}\right)}{A\left(\rm{Cl}\right)} \left\lbrace  1+ \left[\epsilon\left(\rm{Si}\right)- \epsilon\left(\rm{Cl}\right) \right] \cos{\omega t}\right\rbrace.
\label{eqn:SiClperiod} 
\end{eqnarray}
\noindent Here A, B, and $ \epsilon $ are constants, with $\epsilon{}(\rm{Si})\neq \epsilon{}(\mathrm{Cl})$, 
and $ \epsilon\ll1 $ has been assumed. In summary, Fig. \ref{fig:SiClratio} is compatible with a picture in which
any differences between Si/Cl and $1/R^2$ are due to the varying sensitivities of Si and Cl to an external perturbation
coming from the Sun, as seen in Ellis' data. Although a differential temperature effect could in principle also
explain Fig. \ref{fig:SiClratio}, this would require temperature-dependent effects which are much larger than allowed
by the preceding analysis.

\section{Gravitational and Related Influences}
The preceding discussion leads naturally to the question of whether the observed effects in the 
BNL and PTB data could arise from a change in the gravitational potential of the Sun at the 
respective detectors as the Earth passes from perihelion to aphelion and back. In the framework 
of conventional General Relativity (GR) all clocks at a given space-time point run at the same 
rate, irrespective of their internal mechanism. In terms of Eq.~\ref{eqn:SiClperiod}  this would correspond 
to setting  $\epsilon(\rm{Si})=\epsilon(\rm{Cl})$, in which case there would be no fluctuation in the ratio Si/Cl. Since this is 
in obvious disagreement with the BNL data, conventional GR can be ruled out as an explanation 
of the observed effects.

However, this conclusion does not necessarily hold for various alternatives to GR. 
Will \citep{Will74} has shown that in non-metric theories, clocks of different 
construction can in fact behave differently in a gravitational field. Although such a mechanism 
is possible in principle, it fails in practice on quantitative grounds. The Sun's gravitational 
potential $\Phi=GM_{\bigodot}/Rc^{2}$ at the Earth is $\Phi=9\times10^{-9}$, and the fractional change $\Delta\Phi$  between perihelion and aphelion is  $\Delta\Phi=3\times10^{-10}$. 
An effect this small would be undetectable given the BNL and PTB statistics, even if the coefficients 
multiplying $\Delta\Phi$ in a non-metric theory were relatively large. Additionally, constraints on parameters which 
measure deviations from GR are sufficiently restrictive \citep{Will93} to preclude the 
possibility of explaining the BNL and PTB data in terms of any known alternative to GR. 

A class of gravity-related theories which cannot be excluded at present are those in which scalar 
fields are introduced to induce a time variation in fundamental constants such as the fine 
structure constant $\alpha=e^{2}/\hslash{}c$. We have discussed elsewhere \citep{Fischbach09} the possibility that 
although the simplest such theories 
cannot account for the BNL or PTB data, theories in which two or more scalar fields are introduced 
which influence both $\alpha$ and the electron/proton mass ratio (and possibly other quantities) might work.

\section{Environmental Influences During the Solar Flare of 2006 December 13}

We turn in this section to a discussion of possible environmental influences on the NaI detector
system used in \citep{Jenkins09Fapp} during the 2006 December 13 solar flare. This is motivated by
the potential connections between the $^{54}$Mn decay data collected in the time period surrounding the flare,
and the BNL and PTB decay data which are the primary focus of the present paper. As we have already noted in 
the Introduction, the short duration of the solar flare allows us to rule out seasonal temperature, humidity
and pressure influences on the detector system as potential explanations of the observed decrease 
in the $^{54}$Mn counting rate. Hence the focus of this section will be on other possible ``environmental'' 
influences on the detector system during the solar flare, such as neutrons, muons, etc.

Although cosmic ray particles are known to be capable of influencing some 
detectors, there is a generic argument against any mechanism which attributes the drop in the $^{54}$Mn 
count rate to cosmic rays. This is based on the observation that the signal for a $^{54}$Mn decay 
in our system is the detection of the characteristic 834.8 keV photon resulting from the electron-capture process,

\begin{eqnarray}
\nonumber e^{-}+~ ^{54}_{25}\mathrm{Mn}\left( J^{P}=3^{+}\right) \rightarrow\nu_{e}+ ~^{54}_{24}\mathrm{Cr}\left( J^{P}=2^{+}\right) \\
\rightarrow ~^{54}_{24}\mathrm{Cr}\left(g.s.\right) + \gamma \left( 834.8 \mathrm{keV} \right)
\label{54MnDecay} 
\end{eqnarray}

\noindent Hence the 834.8 keV photon which signals the electron-capture process is uniquely associated 
with the specific transition of the excited $ ^{54}_{24}\mathrm{Cr} $ from the excited $ J^{P}=2^{+} $ 
state to the $ J^{P}=0^{+} $ ground state (g.s.). The data points utilized in 
Ref. \citep{Jenkins09Fapp} represent the integral count within the Region of Interest (ROI) set
on the 834.8 keV photon for a four hour live-time. Even if there had been a large change in the cosmic ray 
flux of neutrons or muons during the flare, the chance that the nuclear or electromagnetic interactions 
induced by these particles would have accidentally produced a photon at 834.8 keV, or deposited the energy
equivalent of 834.8 keV within the NaI crystal, is extremely 
small. Moreover, the fact that a deficit of counts appeared in the ROI, and not an increase, indicates that this already 
very remote possibility is highly unlikely. It is also unlikely that counts were selectively removed
from the ROI as a result of changes to the cosmic ray flux. An alternative explanation might be that an 
increase in the cosmic ray flux would generate a large distribution of energy depositions across the energy
spectrum, and thus change the system dead time, leading to a loss of counts within the ROI. Dead time was tracked
as a variable, however, and did not show changes over the course of the flare when compared to the 
counts preceding or following the flare period.
Additionally, inspection of the spectra files also show no significant changes in the counts of channels above or below 
the 834.8 keV $^{54}$Mn peak.

The above argument is bolstered by observations made by other groups during the 2006 December 13 
flare. Although an analysis of data from world-wide neutron monitors by B\"{u}tikofer, et al. and 
Plainaki, et al. \citep{Butikofer09,Plainaki09} exhibited a $\sim$90\% increase in the 1-minute data at 
Oulu and Apatity, the overall neutron flux was still so small as to have a negligible effect on the 
detector system. This is particularly true in light of the fact that the NaI detector is insensitive to neutrons, and
the flux of neutrons was too low to affect the $^{54}$Mn sample directly. 

Similar arguments allow us to exclude a change in the flux of cosmic ray muons resulting from a 
Forbush decrease \citep{Angelov09} as a possible explanation of the flare data. A Forbush decrease 
is a rapid change in the flux of cosmic rays resulting from plasma clouds emitted by the Sun during 
a solar storm \citep{Angelov09}. Evidence for a Forbush decrease during the 2006 December 13 flare 
obtained from the muon telescopes at the Basic Environmental Observatory (BEO) in Bulgaria and 
from Nagoya, Japan are shown in Figs. 5 and 6 of Ref. \citep{Angelov09}, respectively. The BEO data 
exhibited a sharp Forbush decrease beginning at $ \sim $15:30-16:30 UT on 2006 December 14, and reached a 
maximum at 05:00 UT on December 15. The beginning of the Forbush decrease thus 
occurred ~37 hours \textit{after} the dip in the $^{54}$Mn counting rate reported 
in Ref. \citep{Jenkins09Fapp}, which coincided in time with the solar flare at $ \sim $02:40 UT on December 13. 
Additionally, the minimum in the muon count rate associated with the Forbush decrease was observed 
more than 2 days later. The data from the BEO 
are supported by data obtained at Nagoya, which exhibited a similar time-dependence \citep{Angelov09}. 
Given the timing of the Forbush decrease relative to the dip in the $^{54}$Mn counting rate we can 
thus reasonably conclude that the dip was not likely the result of the response of the detector system to 
the Forbush decrease.

The preceding arguments can be further strengthened by noting from Ref. \citep{Jenkins09Fapp} 
that the $^{54}$Mn count rate started to decrease $\sim$1.7 days \textit{before} the flare. It follows that the 
beginning of the decrease in the $^{54}$Mn count rate preceded the beginning of the Forbush 
decrease by more than 70 hours, and hence was not caused by it.

In summary, we can rule out environmental explanations of the correlation between the solar 
flare of 2006 December 13 and the observed dip in our $^{54}$Mn data \citep{Jenkins09Fapp} on 
several grounds. These include, (a) the observation that the solar flare lasted approximately 43 minutes,
which is too short a time for seasonal environmental influence to have affected the detector system. (b) 
the failure of these mechanisms to produce the 
characteristic 834.8 keV photon arising from $^{54}$Mn electron capture, or to account for the $\sim$1.7 day 
precursor signal observed in the $^{54}$Mn decay data; (c) the timing of the Forbush decrease which 
followed, rather than preceded, the dip in the $^{54}$Mn data. Additionally, the observed changes 
in cosmic ray fluxes would have been too small quantitatively to account for the $^{54}$Mn data.

\section{Discussion and Conclusions}

As we noted in the Introduction, there are at present two competing general explanations for the apparent
fluctuations observed in the BNL and PTB data: (a) they arise from the responses of the respective detector
systems to seasonal variations in temperature, pressure and humidity, and possibly other factors such as 
radon buildup. (b) the fluctuations arise from the decay process itself, due to some as yet unknown 
influence possibly originating from the Sun. By modeling the respective detector systems in detail, we 
have shown here that alternative (a) is an unlikely explanation for the observed BNL and PTB data, and 
hence by implication alternative (b) is more likely to be correct.
The resulting inference, that nuclear decay rates are being directly influenced by solar activity is 
further supported by our analysis of $^{54}$Mn decay data acquired during the solar flare 
of 2006 December 13, which we have also analyzed. Our conclusion that the dip in the $^{54}$Mn 
data, which was coincident in time with the flare, was not likely attributable to changes in the cosmic 
ray flux during the flare, further strengthens the case that nuclear decays are being directly 
affected by solar activity. This inference can evidently be checked in a variety of experiments 
such as those described on Ref. \citep{Fischbach09}, some of which are already in progress.

\section{Acknowledgments}

The authors wish to thank D. Alburger and G. Harbottle for making their raw BNL data, apparatus, 
and samples available to us, and for many helpful conversations. We are also deeply grateful 
to H. Schrader for the raw PTB decay data, as well as for invaluable discussions on his 
experiment during our visit to the PTB. We are indebted to our colleagues who have helped us 
in this work, including V. Barnes, J. Buncher, T. Downar, D. Elmore, A. Fentiman, J. Heim, 
M. Jones, A. Karam, D. Krause, A. Longman, J. Mattes, E. Merritt, H. Miser, T. Mohsinally, 
S. Revankar, B. Revis, A. Treacher, J. Schweitzer, and F. Wissman. 
The work of EF was supported in part my the U.S. Department of Energy 
under Contract No. DE-AC02-76ER071428.




\begin{thebibliography}{10}
\expandafter\ifx\csname url\endcsname\relax
  \def\url#1{\texttt{#1}}\fi
\expandafter\ifx\csname urlprefix\endcsname\relax\def\urlprefix{URL }\fi
\expandafter\ifx\csname href\endcsname\relax
  \def\href#1#2{#2} \def\path#1{#1}\fi

\bibitem{Jenkins09Fapp}
J.~H. Jenkins, E.~Fischbach, Perturbation of nuclear decay rates during the
  solar flare of 2006 {D}ecember 13, Astropart. Phys. 31~(6) (2009) 407--411.

\bibitem{Jenkins09Capp}
J.~H. Jenkins, E.~Fischbach, J.~B. Buncher, J.~J. Mattes, D.~E. Krause, J.~T.
  Gruenwald, Evidence of correlations between nuclear decay rates and
  {E}arth-{S}un distance, Astropart. Phys. 32~(1) (2009) 42--46.

\bibitem{Fischbach09}
E.~Fischbach, J.~Buncher, J.~Gruenwald, J.~Jenkins, D.~Krause, J.~Mattes,
  J.~Newport, Time-dependent nuclear decay parameters: New evidence for new
  forces?, Space Sci. Rev. 145~(3) (2009) 285--335.

\bibitem{Alburger86}
D.~E. Alburger, G.~Harbottle, E.~F. Norton, Half-life of $^{32}${S}i, Earth and
  Planet. Sci. Lett. 78~(2-3) (1986) 168--76.

\bibitem{Siegert98}
H.~Siegert, H.~Schrader, U.~Schoetzig, Half-life measurements of europium
  radionuclides and the long-term stability of detectors, Appl. Rad. and Isot.
  49~(9-11) (1998) 1397--1401.

\bibitem{Schrader06}
H.~Schrader, Transmission of $^{226}${R}a data.

\bibitem{Verkho08}
O.~P. Verkhoglyadova, L.~Gang, G.~P. Zank, H.~Qiang, Modeling a mixed {SEP}
  event with the {PATH} model: {D}ecember 13, 2006 1039 (2008) 214--219.

\bibitem{Semkow09}
T.~M. Semkow, D.~K. Haines, S.~E. Beach, B.~J. Kilpatrick, A.~J. Khan,
  K.~O'Brien, Oscillations in radioactive exponential decay, Phys. Lett. B
  675~(5) (2009) 415--419.

\bibitem{Harbottle73}
G.~Harbottle, C.~Koehler, R.~Withnell, A differential counter for the
  determination of small differences in decay rates, Rev. of Sci. Instrum.
  44~(1) (1973) 55--9.

\bibitem{Schrader07}
H.~Schrader, Ionization chambers, Metrologia 44~(4) (2007) 53--66.

\bibitem{Knoll00}
G.~F. Knoll, Radiation detection and measurement, 3rd Edition, John Wiley \&
  Sons, New York, 2000.

\bibitem{Found20}
C.~G. Found, Ionization potentials of argon, nitrogen, carbon monoxide, helium,
  hydrogen and mercury and iodine vapors, Phys. Rev. 16~(1) (1920) 41.

\bibitem{Okabe87}
S.~Okabe, T.~Nishikawa, M.~Aoki, M.~Yamada, Looping variation observed in
  environmental gamma-ray measurement due to atmospheric radon daughters A255
  (1987) 371--3.

\bibitem{Nishikawa88}
T.~Nishikawa, S.~Okabe, M.~Aoki, Seasonal variation of radon daughters
  concentrations in the atmosphere and in precipitation at the {J}apanese coast
  of the sea of {J}apan 24 (1988) 93--5.

\bibitem{Miles88}
J.~C.~H. Miles, R.~A. Algar, Variations in radon-222 concentrations, Jour. of
  Radiologic. Prot. 8~(2) (1988) 103--105.

\bibitem{Gaso94}
M.~I. Gaso, M.~L. Cervantes, N.~Segovia, V.~H. Espindola, Atmospheric radon
  concentration levels, Radiat. Meas. 23~(1) (1994) 225--30.

\bibitem{Thorne03}
M.~C. Thorne, Background radiation: natural and man-made, Jour. of Radiologic.
  Prot. 23~(1) (2003) 29--42.

\bibitem{Abbady04}
A.~Abbady, A.~G.~E. Abbady, R.~Michel, Indoor radon measurement with the lucas
  cell technique, Appl. Rad. and Isot. 61~(6) (2004) 1469--1475.

\bibitem{Wissmann06}
F.~Wissmann, Variations observed in environmental radiation at ground level,
  Radiat. Protect. Dosimetry 118~(1) (2006) 3--10, 10.1093/rpd/nci317.

\bibitem{Szegvary09}
T.~Szegvary, F.~Conen, P.~Ciais, European $^{222}${R}n inventory for applied
  atmospheric studies, Atmospheric Environment 43~(8) (2009) 1536--1539.

\bibitem{Schrader00}
H.~Schrader, Calibration and consistency of results of an ionization-chamber
  secondary standard measuring system for activity, Appl. Rad. Isot. 52~(3)
  (2000) 325--334.

\bibitem{Artna-Cohen96}
A.~Artna-Cohen96, Nuclear data sheets for {E}u-152, Nucl. Data Sheets 79~(1).

\bibitem{Singh06}
B.~Singh, Nuclear data sheets for 32si, Tech. rep. (2006).

\bibitem{Endt99P32}
P.~M. Endt, R.~B. Firestone, Nuclear data sheets for $^{32}${P}, Tech. rep.
  (1999).

\bibitem{Endt99Cl36}
P.~M. Endt, R.~B. Firestone, Nuclear data sheets for $^{36}${C}l, Tech. rep.
  (1999).

\bibitem{Cumming83}
J.~B. Cumming, Assay of silicon-32 by liquid scintillation counting, Radiochem.
  and Radioanalyt. Lett. 58~(5-6) (1983) 297--305.

\bibitem{Giacomo82}
P.~Giacomo, Equation for the determination of the density of moist air (1981),
  Metrologia 18~(3) (1982) 171--171.

\bibitem{Davis92}
R.~S. Davis, Equation for the determination of the density of moist air
  (1981/91), Metrologia 29~(1) (1992) 67--70.

\bibitem{Picard08}
A.~Picard, R.~S. Davis, M.~Glaser, K.~Fujii, Revised formula for the density of
  moist air ({CIPM}-2007), Metrologia 45~(2) (2008) 149--155.

\bibitem{Katz52}
L.~Katz, A.~S. Penfold, Range-energy relations for electrons and the
  determination of beta-ray end-point energies by absorption, Rev. Mod. Phys.
  24 (1952) 28--44.

\bibitem{Pelowitz08}
Los Alamos National Laboratory, MCNPX User's Manual, Version 2.6.0,
  la-cp-07-1473 Edition (April 2008).

\bibitem{NCDC}
NCDC, \href{http://www.ncdc.noaa.gov/oa/land.html}{National climatic data
  center: Land-based data} (June 2009).
\newline\urlprefix\url{http://www.ncdc.noaa.gov/oa/land.html}

\bibitem{Ellis90}
K.~J. Ellis, The effective half-life of a broad beam $^{238}${P}u/{B}e total
  body neutron irradiator, Phys. Med. Biol. 35~(8) (1990) 1079.

\bibitem{Will74}
C.~M. Will, Gravitational red-shift measurements as tests of non-metric
  theories of gravity, Phys. Rev. D 10~(8) (1974) 2330.

\bibitem{Will93}
C.~M. Will, Theory and experiment in gravitational physics, rev. Edition,
  Cambridge University Press, Cambridge [England] ; New York, NY, USA, 1993.

\bibitem{Butikofer09}
R.~B\"{u}tikofer, E.~O. Fl\"{u}ckiger, L.~Desorgher, M.~R. Moser, B.~Pirard,
  The solar cosmic ray ground-level enhancements on 20 {J}anuary 2005 and 13
  {D}ecember 2006, Adv. Space Res. 43~(4) (2009) 499.

\bibitem{Plainaki09}
C.~Plainaki, H.~Mavromichalaki, A.~Belov, E.~Eroshenko, V.~Yanke, Modeling the
  solar cosmic ray event of 13 {D}ecember 2006 using ground level neutron
  monitor data, Adv. in Space Res. 43~(4) (2009) 474.

\bibitem{Angelov09}
I.~Angelov, E.~Malamova, J.~Stamenov, The {F}orbush decrease after the {GLE} on
  13 {D}ecember 2006 detected by the muon telescope at {BEO} - {M}oussala, Adv.
  Space Res. 43~(4) (2009) 504.

\end{thebibliography}
\end{document}